\documentclass[5p,review,times,2022]{elsarticle}

\usepackage{adjustbox}

\usepackage{array}
\usepackage{graphicx}
\usepackage{hyperref}
\usepackage{booktabs}
\usepackage{amsmath}
\usepackage{amssymb}
\newcommand{\norm}[1]{\left\lVert#1\right\rVert}
\usepackage{float}
\usepackage[dvipsnames]{xcolor}
\usepackage{caption}
\usepackage{subcaption}

\begin{document}

\title{Capacity Analysis of Intersections When CAVs Crossing in a Collaborative and Lane-Free Order}

\author[]{Mahdi Amouzadi}
\ead{m.amouzadi@sussex.ac.uk}

\author[]{Mobolaji Olawumi}

\author[]{Arash M. Dizqah}
\ead{a.m.dizqah@sussex.ac.uk}

\address{Smart Vehicles Control Laboratory (SVeCLab), University of Sussex, Brighton BN1 9RH, UK}


\begin{abstract}
Connected and autonomous vehicles (CAVs) improve the throughput of intersections by crossing in a lane-free order as compared to the signalised crossing of human drivers. However, it is challenging to quantify such an improvement because the available frameworks to analyse the capacity (i.e., the maximum throughput) of the conventional intersections does not apply to the lane-free ones. This paper proposes a novel theoretical framework to numerically simulate and compare the capacity of lane-free and conventional intersections. The results show that the maximum number of vehicles passing through a lane-free intersection is up to seven times more than a signalised intersection managed by the state-of-the-art max-pressure and Webster algorithms. A sensitivity analysis shows that, in contrast to the signalised intersections, the capacity of the lane-free intersections improves by an increase in initial speed, the maximum permissible speed and acceleration of vehicles.
\end{abstract}

\begin{keyword}
intersection throughput; traffic management; connected and autonomous vehicles; signal free intersection; signalised intersection
\end{keyword}


\maketitle


%

\section{Introduction}

Capacity analysis of intersections is essential for the management of traffic and planning transport systems. Unlike traditional human-driven vehicles (HVs) which are restricted to travel within the road lanes, connected and autonomous vehicles (CAVs) enable a lane-free crossing through intersections. There are extensive prior works to characterise the capacity of intersections for HVs (e.g. \cite{national2000highway, farivar2015modeling,wu2016capacity}), however, such analysis for CAVs in a lane-free order is still an open research topic.

Whilst human reaction is the dominant factor to measure capacity of intersections with HVs, CAVs are driverless vehicles where the reaction time is not meaningful. The measures of capacity for HVs crossing both signalised and unsignalised (two-way stop-controlled and all-way stop-controlled) intersections are extensively discussed in Highway Capacity Manual \cite{national2000highway}. The manual introduces a measure to quantify the capacity of the unsignalised two-way and all-way stop-controlled intersections based on, respectively, gap acceptance and queuing theories. Meanwhile, it is recommended in \cite{national2000highway} to calculate the capacity of the signalised intersections as the saturation flow rate times the green time ratio. All of these measures assume that headway of each HV in the queue of lanes is known to be around 1.9~s. This assumption makes these measures inappropriate for lane-free intersections where the headway of CAVs is much smaller and almost the same for all the vehicles in the queue \cite{wu2021junction}.

To evaluate capacity of the intersections with CAVs, the authors in \cite{yu2019managing,wu2021junction,li2019method} employed the same measure that is defined in \cite{national2000highway} for the unsignalised intersections, though with a new headway definition for CAVs. In \cite{yu2019managing}, intersections are assumed as service providers and CAVs headway is redefined as service time (i.e., crossing time) which is derived by applying queuing theory. The service time is based on the safety time gap of CAVs approaching the intersection from the same stream and from the conflicting streams. A similar work is proposed in \cite{wu2021junction} that employs the M/G/1 queue model to drive a formula for the capacity of the intersections. This model assumes that the intersection capacity is equivalent to the service rate of vehicles. Finally, the authors in \cite{li2019method} reformulated the capacity measure of the unsignalised two-way stop-controlled intersections to use the critical gap and follow-up time of CAVs instead of the ones of HVs. The measures provided by these researchers are effective to evaluate capacity of the intersections when CAVs drive through a restricted set of lanes, however, are not applicable to the lane-free intersections. Hence, there is a need for a measure to quantify the capacity for the lane-free crossing of CAVs through intersections. Moreover, the measure must be uniformly applicable to both the CAVs crossing lane-free intersections and HVs crossing signalised intersections to make it possible to quantify the capacity improvements of intersections by CAVs.

It is also important to note that CAVs are heterogenous in terms of their control strategy \cite{yu2019managing} and any measure to quantify capacity of intersections must be independent of the performance of these strategies. However, the majority of the above-mentioned research measure the capacity of intersections that are controlled by a reservation-based strategy.

\begin{figure*}[h]
    \centering
	\includegraphics[scale=0.55]{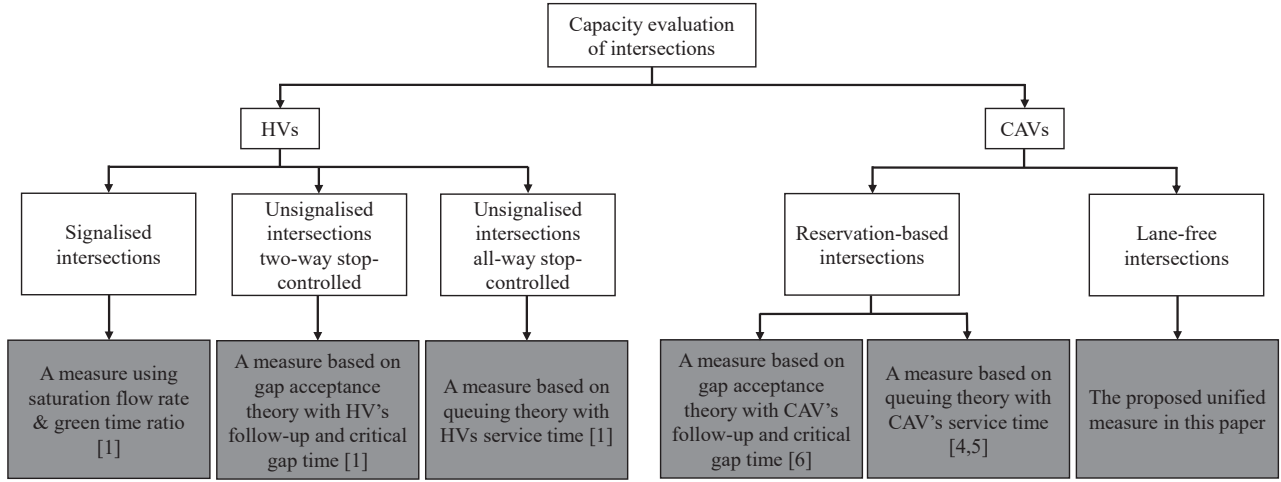}
	\caption{The measures for capacity evaluation of intersections (grey boxes) with both the human-driven vehicles (HVs) and connected and autonomous vehicles (CAVs).}
	\label{block}
	\vspace{-16pt}
\end{figure*}

The first type of reservation strategies is called intersection-reservation where the controller reserves the whole intersection for one CAV at a time. The authors in \cite{pan2020optimal} formulated an optimal control problem (OCP) to minimise the crossing time of CAVs while reserving the whole intersection to avoid collisions. The references \cite{xu2018distributed} and \cite{di2019design} introduce an intersection crossing algorithm where CAVs are placed into a virtual platoon based on their distance to the centre of the intersection. The algorithm reserves the whole intersection for each CAV in the platoon to pass through without collision. Generally speaking, reservation of the whole intersection reduces the capacity of the intersection. 

The second type is a conflict-point-reservation strategy that reduces the area of reservation to just a few conflicting points. The authors in \cite{wang2020cooperative,mirheli2019consensus} designed optimisation-based algorithms to realise this type of reservation algorithms. A similar work is proposed in \cite{zhang2021priority}, where a constraint is added to the optimisation problem for each conflict point to limit the maximum number of crossing vehicles at any time to one. Even though the reservation-based strategies improve the capacity of intersections by nullifying the stop-and-go requirement of the conventional signalised intersections, yet vehicles must follow a set of predefined paths and are not able to fully utilise the intersection area by lane-free manoeuvres. 

Alternatively, the authors in \cite{li2018near} developed an OCP to formulate the lane-free crossing problem of intersections. The objective of the developed OCP is to minimise the crossing time of CAVs and therefore the algorithm generates time-optimal trajectories for each CAV. However, the proposed OCP contains a set of highly non-convex constraints to represent the collision avoidance criteria which makes it difficult to solve online. To resolve this issue, Li et al. \cite{li2018near} splits the non-convex formulation into two stages. At stage one which is solved online, CAVs make a standard multi-lane formation by moving to pre-defined positions of each lane. At stage two, the controller determines the crossing scenario based on destinations of CAVs in the formation (depending number of lanes the number of possible scenarios could be significantly high). The controller, then, fetches the optimal solution of the lane-free crossing of the CAVs of this particular scenario from a look-up table and enforces the CAVs to follow the pre-defined trajectories. The solution of the non-convex OCP for each scenario is already calculated offline and stored in the look-up table. The approach is scientifically interesting but it is not practical because, for example, it takes around 356 years to solve the non-convex OCPs for all possible scenarios using the state-of-the-art processors when there is 24 CAVs \cite{li2018near}. 

In a more recent study, Li et al \cite{li2020autonomous} changed the minimum-time OCP of \cite{li2018near} to a feasibility problem to make the non-convex formulation tractable. However, this results in a sub-optimal solution. The authors in \cite{https://doi.org/10.48550/arxiv.2204.03550} resolved the previous issues by using dual problem theory to convexify the non-convex constraints that avoid CAVs colliding with each other and with road boundaries. This work generates time-optimal trajectories of CAVs passing through intersections in a lane-free order and shows that such a lane-free crossing reduces the travelling time by up to 65\% as compared to the state-of-the-art reservation-based method proposed in \cite{malikopoulos2021optimal}. 

Fig. \ref{block} summarises different measures that are proposed by prior works to calculate the capacity of intersections for both HVs and CAVs.

This paper addresses the above-mentioned gap, i.e. the lack of a measure to quantify the capacity of lane-free intersections regardless of control strategies, by the following contributions to the knowledge:

\begin{itemize}
    \item A novel framework to uniformly evaluate the capacity of both the lane-free and signalised intersections with, respectively, CAVs and HVs crossing through. The framework consists of a novel measure of capacity along with an algorithm to calculate this measure for both cases.
    \item Assessment of the capacity improvement by the lane-free crossing of CAVs as compared to the conventional signalised crossing of HVs through intersections.
    \item A sensitivity analysis of the capacity and crossing time of intersections with respect to the maximum speed, maximum acceleration/deceleration, initial speed and the number of the crossing vehicles. The analysis includes both the lane-free and signalised crossing.
\end{itemize}

The remainder of this paper is structured as follows: Section \ref{sys} provides the theoretical scheme that represents the lane-free intersections in this study. Section \ref{framework} introduces the proposed framework including a novel measure and the algorithms to calculate the measure for the lane-free and signalised intersections. The capacity improvement of lane-free intersections as compared to signalised intersections is demonstrated in Section \ref{improvement}. A sensitivity analysis for both the lane-free and signalised intersections with respect to different values of speed and acceleration is presented in Section \ref{sensitivity} followed by a conclusion in Section \ref{conclusion}.

\section{System Description}\label{sys}
\begin{figure}[t]
    \centering
	\includegraphics[scale=0.4]{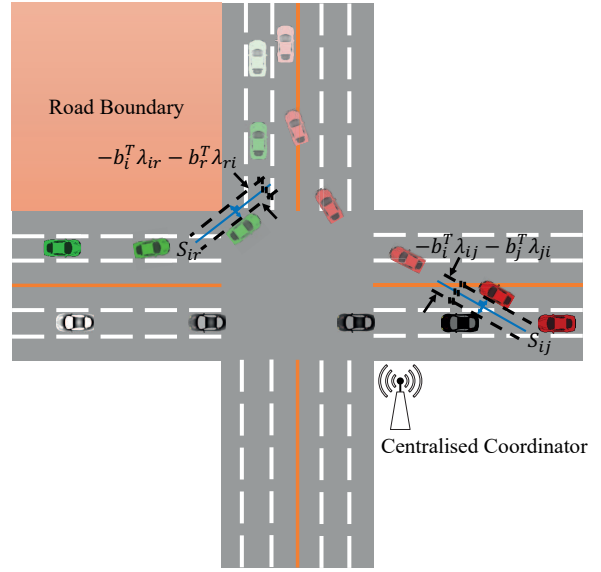}
	\caption{A lane-free and signal-free intersection. $S_{ij}$ and $S_{ir}$ are the separating hyperplanes between, respectively, two CAVs and a CAV and road boundary. $-b_{i}^\top\lambda_{ij} - b_{j}^\top\lambda_{ji}$ and $-b_{i}^\top\lambda_{ir} - b_{r}^\top\lambda_{ri}$ are distances which are formulated from the equivalent dual problem of the obstacle avoidance constraints. The dashed lines are the supporting hyperplanes.}
	\label{Intersection_layout}
	\vspace{-10pt}
\end{figure}

Fig. \ref{Intersection_layout} shows an example of a lane-free and signal-free intersection. The intersection is composed of four approaches, each approach with an incoming and an outgoing lane. The colour brightness of the three crossing CAVs changes from solid at the starting point to the most transparent at the destination. Lane-free intersections allow CAVs to change their lanes at any point of journey that helps to travel faster. For instance, the red CAV in Fig. \ref{Intersection_layout} takes over the black CAV by travelling to the opposite lane. Moreover, this intersection has no traffic light and CAVs collaborate to cross safe and fast. This study, however, focuses on the global optimum crossing of CAVs and therefore assumes that there is a centralised coordinator for this purpose that is placed at the intersection.

The formulated collision avoidance constraints for the red and black CAVs after applying the dual problem theory \cite{https://doi.org/10.48550/arxiv.2204.03550} are also shown in Fig. \ref{Intersection_layout}. The expression $-b_{i}^\top\lambda_{ij} - b_{j}^\top\lambda_{ji}$ represents the dual form of the distance between the two CAVs where $S_{ij}$ is the separating hyperplane placed between them. Similarly, the expression $-b_{i}^\top\lambda_{ir} - b_{r}^\top\lambda_{ri}$ represents the dual form of the distance between the green CAV and the highlighted road boundary and $S_{ir}$ is the separating hyperplane. For more details on the convexification of collision avoidance constraints, refer to \cite{https://doi.org/10.48550/arxiv.2204.03550}.

It is worth noting that capacity of intersections is defined in this study as the maximum number of vehicles that can continuously cross the intersection within a unit of the time and under predefined conditions (e.g, intersection geometry, roadway and distribution of traffic) \cite{yu2019managing,lioris2017platoons}. This definition is equivalent to the service rate in queuing theory and it is not the same as the one provided in HCM \cite{national2000highway}. 

\section{A Novel Framework to Quantify Capacity of Intersections}\label{framework}

Conventionally, capacity of intersections (both the signalised and unsignalised) are measured using a set of collected data from either real-time observation of vehicles \cite{national2000highway} or running a micro-simulation \cite{chaudhry2009capacity,chauhan2022microscopic}. For example, the capacity of each lane of an unsignalised all-way stop-controlled intersection is measured when the degree of utilisation (DoU) of the lane reaches one. DoU represents the fraction of capacity being used by vehicles and is defined as follows \cite{national2000highway}: 
\begin{equation}\label{cap_HCM}
    x = \frac{v h_d}{3600}\qquad 
\end{equation}
where $x$ denotes the degree of utilisation, $v$ refers to flow rate (throughput) $(veh/h)$ of the lane and $h_d$ is the departure headway $(s)$ that is calculated as a stochastically weighted average of the saturation headway of all combinations of possible degrees of conflict and number of crossing vehicles. 

The reference \cite{national2000highway} proposes an iterative algorithm to calculate the value of $x$ and $h_d$ for any given $v$ based on some identified values from the available large set of real data. The capacity is then determined by gradually increasing the throughput ($v$) until the calculated value $x$ reaches to one where the capacity equals the throughput $v$. 

However, such real-time data are not available for CAVs (especially for lane-free crossing of CAVs) because of the lack of real infrastructure or realistic simulators that consider enough number of heterogeneous CAVs crossing an intersection. 
 \subsection{The proposed novel measure of capacity of intersections}\label{sssec:num1}
Intersections can host limited number of vehicles at the same time and if the intersection capacity exceeds, either the waiting time of crossing vehicles will significantly increase or collisions will happen. Therefore, to evaluate the capacity of intersections a suitable measure must consider the maximum number of vehicles and the time that it takes for those vehicles to pass through the intersection without any collisions. In effect, the following measure is proposed to calculate the capacity of any intersection:

\begin{equation}\label{capacity_proposed}
    C = max\big\{\frac{3600*N}{T}\big\}
\end{equation}
where $C$ is the capacity $(veh/h)$ of the intersection, $N$ denotes the number of crossing HVs/CAVs $(veh)$ and $T$ represents the time $(s)$ that takes for those vehicles to fully cross the intersection. 

It is already shown in \cite{https://doi.org/10.48550/arxiv.2204.03550} that the minimum crossing time $T_{min}$ of CAVs through a lane-free intersection is almost fixed to a constant value regardless of the number of CAVs. Hence, if one can calculate $T_{min}$, then the capacity of lane-free intersections is calculated by finding the number of crossing CAVs $N_{max}$ just before a collision occurs. Therefore, the capacity of a lane-free intersection is calculated as follows:

\begin{equation}\label{capacity_CAV}
    C_{lf} = \frac{3600*N_{max}}{T_{min}}
\end{equation}

For conventional intersections, however, the crossing time changes by increasing the number of crossing vehicles and hence one needs a simulator to gradually increase the number of crossing vehicles and measure the crossing time and the capacity will be calculated as in (\ref{capacity_proposed}) at the point where the throughput starts declining. It is obvious that by increasing the number of vehicles crossing signalised intersections collisions would not happen, however, the crossing time significantly increases which shows the capacity is reached.

The next sections present the algorithms to measure $T_{min}$ and $N_{max}$ for lane-free intersections, as well as the maximum throughput of the signalised intersections.

\subsection{Calculation of $T_{min}$ and $N_{max}$ for lane-free intersections}\label{sssec:num2}
The time-optimal crossing algorithm proposed in \cite{https://doi.org/10.48550/arxiv.2204.03550} for the lane-free intersections is employed to calculate the minimum crossing time $T_{min}$ and the maximum number of CAVs $N_{max}$ that can cross the intersection within $T_{min}$ without collision. The work formulates a centralised convexified OCP to generate global optimum crossing time of CAVs. The formulated time-optimal OCP is solved for gradually increasing number of CAVs to determine the $N_{max}$. For the sake of completeness of this paper, a summary of the algorithm, including the considered dynamics of the vehicles, collision avoidance constraints and the resulting OCP, is also explained here.

\vspace{3pt}
\subsubsection{Modelling of CAVs and road boundaries}
CAVs are represented in this study with their two degree-of-freedom (DoF) bicycle models \cite{milliken1995race}. The lateral motion of the vehicle is modelled with two DoFs which are the sideslip angle $\beta_{i}$ and the yaw rate $r_{i}$. The longitudinal motion, on the other hand, is modelled by the longitudinal speed $V_{i}$ of the vehicle as an additional DoF. The following differential equations present the vehicle model of $\text{CAV}_{i}$ where $i \in \{1..N\}$ and $N$ is the total number of CAVs:

\vspace{-10pt}
\begin{align}
    \frac{d}{dt} \begin{bmatrix}\notag
    r_{i}\\[0.1cm]
    \beta_{i}\\[0.1cm]
    V_{i}\\[0.1cm]
    x_{i}\\[0.1cm]
    y_{i}\\[0.1cm]
    \theta_{i}\\[0.1cm]
    \end{bmatrix}(t)
    =
    &\begin{bmatrix}
    \frac{N_{r}}{I_{z}} \cdot r_{i}(t) + \frac{N_{\beta}}{I_{z}} \cdot \beta_{i}(t)\\[0.1cm]
    (\frac{Y_{r}}{m \cdot V_{i}(t)} - 1) \cdot r_{i}(t) + \frac{Y_{\beta}}{m\cdot V_{i}(t)} \cdot \beta_{i}(t) \\[0.1cm]
     0\\[0.1cm]
    V_{i}(t) \cdot cos\theta_{i}(t)\\[0.1cm]
    V_{i}(t) \cdot sin\theta_{i}(t)\\[0.1cm]
    r_{i}(t)\\[0.1cm]
    \end{bmatrix}
    +\\ 
    &\qquad \qquad\begin{bmatrix}
    0 & \frac{N_{\delta}}{I_{z}} \\ 
    0 & \frac{Y_{\delta}}{m \cdot V_{i}(t)}  \\ 
    1 & 0 \\
    0 & 0 \\
    0 & 0 \\ 
    0 & 0 \\ 
    \end{bmatrix}
    \begin{bmatrix}
    a_{i}\\
    \delta_{i}
    \end{bmatrix}(t)
    \label{vehilce_model}
\end{align}
where the control inputs and states of $\text{CAV}_{i}$ are presented as $\textbf{u} =[a_{i}, \delta_{i}]^T$ and $\textbf{x} = [r_{i},\beta_{i},V_{i},x_{i},y_{i},\theta_{i}]^T$, respectively. The pose of $\text{CAV}_{i}$ at time $t$ is denoted as $z_i(t) = [x_i(t), y_i(t), \theta_i(t)]^T$. $\delta_{i}(t)$ and $a_{i}(t)$ are the wheel steering angle ($rad$) and acceleration ($m/s^2$) of $\text{CAV}_{i}$. The constants $m$ and $I_z$ represent, respectively, the mass ($kg$) of the vehicle and its moment of inertia ($kg.m^2$) around axis $z$. The vehicle parameters $Y_r$, $Y_\beta$, $Y_\delta$, $N_r$, $N_\beta$ and $N_\delta$ are calculated as in \cite{milliken1995race}.

To ensure CAVs drive within their admissible range, the following constraints are imposed for each $\text{CAV}_{i}$:

 \vspace{-10pt}
\begin{subequations}\label{Limits}
    \begin{align}
        \underline{V} \leq V_{i}(t)\leq \bar{V},\\
        \underline{a} \leq \rvert a_{i}(t) \rvert \leq \bar{a},\\
        \underline{\delta} \leq \rvert \delta_{i}(t) \rvert \leq \bar{\delta},\\
        \underline{r} \leq \rvert r_{i}(t) \rvert \leq \bar{r},\\
        \underline{\beta} \leq \rvert \beta_{i}(t) \rvert \leq \bar{\beta}.
     \end{align}
\end{subequations}
where $\underbar{.}$ and $\overline{.}$ are the lower and upper boundaries respectively.

Each $\text{CAV}_{i}$ is presented as a rectangular polytope $\beta_{i}$ where it is composed of intersection of half-space linear inequality $A_{i}X \le b_{i}$ where $X = [x,y]^T$ is a Cartesian point. 

Road boundaries are also modelled with convex polytopic sets $O_{r}$ where $r \in \{1..N_{r}\}$ and $N_{r}$ denotes the total number of road boundaries which is 4 for a four-legged intersection.

\vspace{3pt}
\subsubsection{Time-optimal OCP for CAVs crossing a lane-free intersection}
The OCP that minimises the crossing time while avoiding collisions is formulated as follows \cite{https://doi.org/10.48550/arxiv.2204.03550}:

 \vspace{-10pt}
\begin{subequations} \label{OCP}
\begin{alignat}{10}
&\{a_i(.),\delta_i(.)\}^* = \\ &\quad \quad\text{arg}\:\underset{\substack{t_{f},a_i(.),\delta_i(.)\\\lambda_{ij},\lambda_{ji},s_{ij},\\\lambda_{ri},\lambda_{ir},s_{ir}}}{\text{minimise}}  J(\textbf{z}_1(.),..,\textbf{z}_N(.))\notag \\
 & \quad \quad \text{s.t.} \quad (\ref{vehilce_model}), (\ref{Limits}),\\
 &\qquad \qquad    -b_{i}(\textbf{z}_{i}(t))^\top\lambda_{ij}(t)-b_{j}(\textbf{z}_{j}(t))^\top \lambda_{ji}(t)\ge d_{min} \label{CA_A} \\
 &\qquad \qquad   A_{i}(\textbf{z}_{i}(t))^\top \lambda_{ij}(t)+s_{ij}(t)=0 \label{CA_B}\\
 &\qquad \qquad    A_{j}(\textbf{z}_{j}(t))^\top \lambda_{ji}(t)-s_{ij}(t)=0 \label{CA_C} \\
  & \qquad \qquad   -b_{i}(\textbf{z}_{i}(t))^\top\lambda_{ir}(t)-b_{r}^\top \lambda_{ri}(t)\ge d_{rmin} \label{RA_A}\\
 &\qquad \qquad    A_{i}(\textbf{z}_{i}(t))^\top \lambda_{ir}(t)+s_{ir}(t)=0 \label{RA_B}\\
 &\qquad \qquad    A_{r}^\top \lambda_{ri}(t)+s_{ir}(t)=0 \label{RA_C}\\
 &\qquad \qquad    \lambda_{ij}(t),\;\lambda_{ji}(t),\;\lambda_{ir}(t),\;\lambda_{ri}(t)\geq 0, \\ 
 & \qquad \qquad   \norm{s_{ij}(t)}_{2}\leq 1,\;\norm{s_{ir}(t)}_{2}\leq1,\\
 &  \qquad \qquad  \textbf{z}_i(t_{0}) = z_{i,0},\:\textbf{z}_i(t_{f}) = z_{i,f},\\
 &  \qquad \qquad   \forall i\neq j \in \{1..N\}, \forall r \in \{1..N_r\}. \notag
\end{alignat}
\end{subequations}
where $\lambda_{ij},\lambda_{ji},s_{ij},\lambda_{ri},\lambda_{ir},s_{ir}$ are dual variables and the subscripts $i$, $j$ and $r$ refer to, respectively, $\text{CAV}_{i}$, $\text{CAV}_{j}$ and $r$th road boundary. $A_{i}$ and $b_{i}$ represent the size and location of $\text{CAV}_{i}$ which are functions of the $\text{CAV}_{i}$'s pose $z_{i}(t)$. (\ref{CA_A})-(\ref{RA_C}) constrain CAVs to avoid collisions with each other and with road boundaries \cite{https://doi.org/10.48550/arxiv.2204.03550}. 

Problem (\ref{OCP}) is nonlinear and is solved using CasADI \cite{Andersson2019} and IPOPT \cite{wachter2006implementation} for any given number $N$ and initial location of CAVs over an unknown final time $t_f$. The solution consists of the final time $t_{f}$ and the optimal trajectories of control inputs $a_i(.)^*$ and $\delta_i(.)^*$ for each $\text{CAV}_i$ over $t\in[t_0,t_f]$. 

The minimum crossing time of CAVs through a given lane-free intersection is fixed to a constant number \cite{https://doi.org/10.48550/arxiv.2204.03550} and, as mentioned before, one solves the problem (\ref{OCP}) for a gradually increasing number of CAVs (until a collision happens) to determine the capacity of the intersection. Then that maximum number of CAV without collision and the corresponding crossing time are used in Eq. \ref{capacity_CAV} to find the capacity of lane-free intersections. 

\subsection{Measurement of the capacity of signalised intersections}\label{sssec:num3}
This study employs the Webster \cite{webster1958traffic} and max-pressure \cite{varaiya2013max} algorithms as the adaptive traffic controllers of the signalised intersections. Webster in \cite{webster1958traffic} derived a formulation that calculates the cycle length of traffic lights. The derived cycle length is used to find the green time of each phase to allow vehicles to cross the intersection. Similarly, the max-pressure algorithm calculates the signal timings, however, the green time of each phase is calculated based on the number of vehicles in the incoming and outgoing lanes \cite{varaiya2013max}.

Whilst Webster is a well-known algorithm for timing control of traffic lights, it is already shown that the max-pressure algorithm yields the lowest travelling time, queues length and crossing delays among all the state-of-the-art controllers, including the algorithms based on the self organising \cite{gershenson2004self,cools2013self}, deep Q-network \cite{lillicrap2015continuous}, deep deterministic policy gradient \cite{casas2017deep} and Webster \cite{webster1958traffic} methods \cite{genders2019open}.

To calculate the capacity of signalised intersections, this study simulates both the max-pressure and Webster algorithms in SUMO for gradually increasing number of HVs based on the works in \cite{genders2019open}. As the number of vehicles increases, throughput of the intersection also increases until reaching its capacity (i.e., $C$ in (\ref{capacity_CAV})) where adding more vehicles causes a drop in throughput due to a sharp rise in the crossing time.

\section{Capacity of the Lane-Free Intersections as Compared to the Signalised Intersections} \label{improvement}
The uniform definition of capacity as in (\ref{capacity_proposed}) makes it possible to determine the amount of traffic improvement by the lane-free crossing of intersections as compared to the signalised crossing. The studied intersection is as in Fig. \ref{Intersection_layout} where the length and width of each lane are, respectively, $50~m$ and $5~m$. The proposed lane-free algorithm in Section \ref{sssec:num2} and the signalised max-pressure and Webster algorithms in Section \ref{sssec:num3} are applied to calculate, respectively, the lane-free and signalised capacities of the intersection for an increasing number of crossing vehicles from 3 to 30. The crossing scenarios for each number of crossing vehicles consists of at least one turning maneuver and is the same for both cases. Table \ref{settings} shows critical parameters that are used in both intersections. A Linux Ubuntu server with a 3.7 GHz Intel core i7 and 32 GB memory is used for all the required calculations. The lane-free algorithm is implemented in Matlab and the signalised algorithms are implemented in SUMO with the help of TraCI.

\renewcommand{\arraystretch}{1.2}
\begin{table}[t]
    \centering
    \caption{Main parameters of the proposed algorithms and their values} 
    \label{settings} 
    \begin{tabular}{l l c}
    \toprule
    \textbf{Parameter(s)} & \textbf{Unit}  & \textbf{Value(s)}\\ [0.5ex] 
    \midrule
    Maximum speed& \multicolumn{1}{m{2.5cm}}{ ($m/s$)} & 25  \\[0.5ex]
    Maximum acceleration & \multicolumn{1}{m{2.5cm}}{($m/s^2$)} & 3 \\[0.5ex]
    Initial speed & \multicolumn{1}{m{2.5cm}}{($m/s$)} & 10 \\[0.5ex]
    Prediction horizon & \multicolumn{1}{m{2.5cm}}{($\times$ \text{sampling times})} & 15 \\[0.5ex]
    Safe margin between CAVs & \multicolumn{1}{m{2.5cm}}{($m$)} & 0.1 \\[0.5ex]
    \bottomrule
    \end{tabular}
\end{table}

\begin{figure}[t]
    \centering
	\includegraphics[scale=0.26]{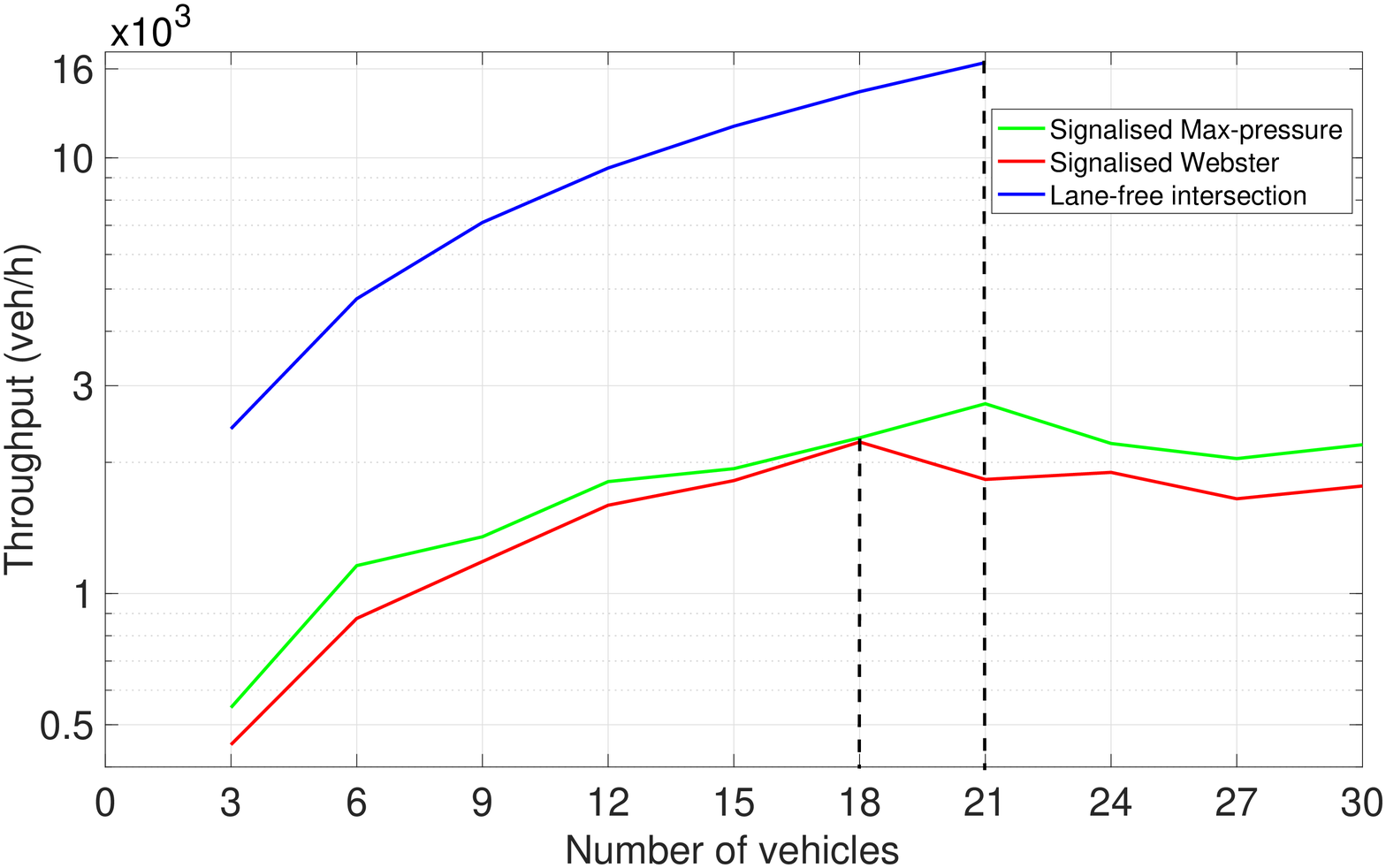}
	\caption{Throughput at different number of vehicles for lane-free and signalised intersections}
	\label{num}
	\vspace{-16pt}
\end{figure}

\begin{figure*}[t]
    \centering
    \begin{subfigure}[t]{0.49\textwidth}
         \centering
         \includegraphics[scale=0.23]{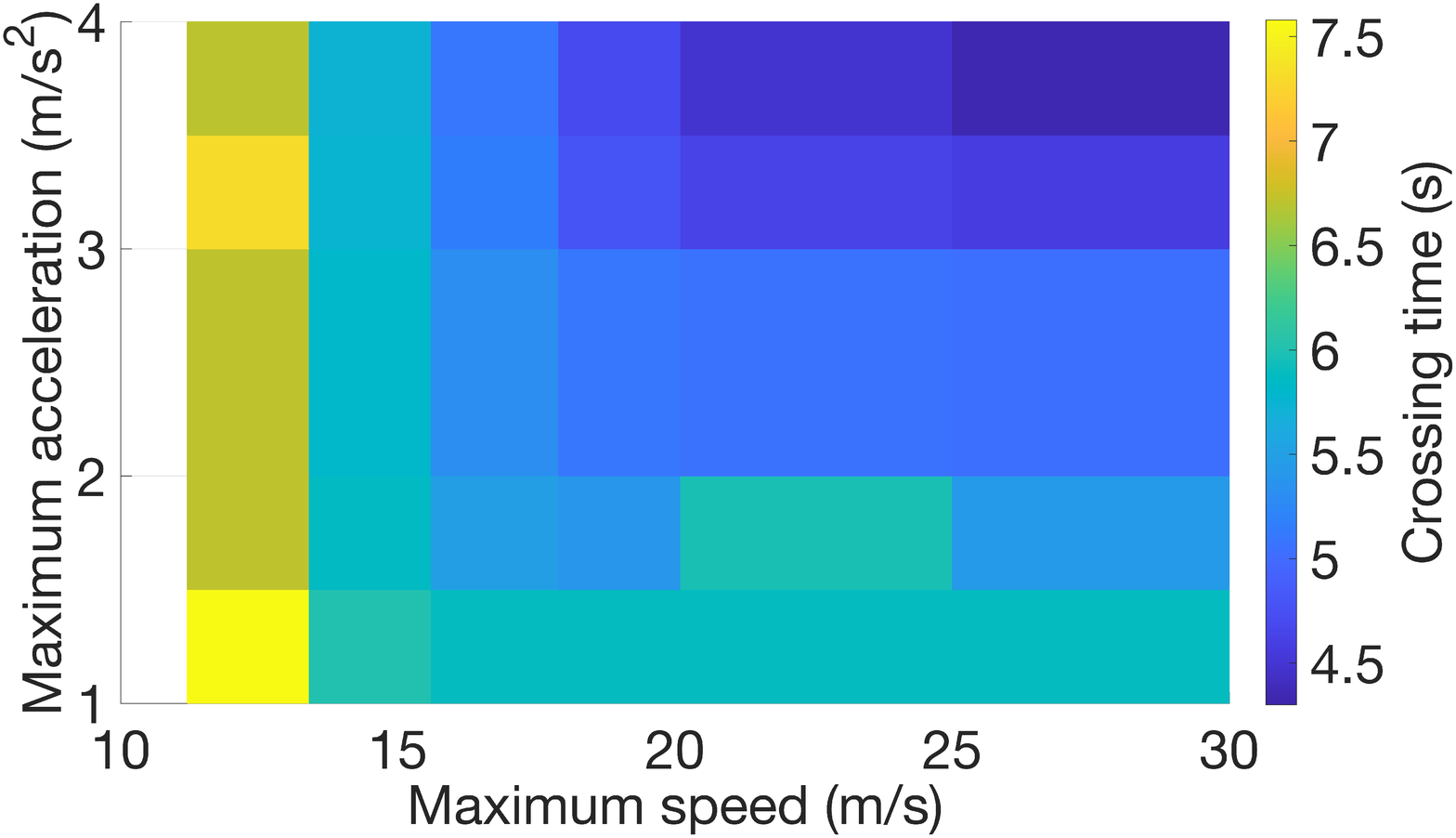}
         \caption{}
         \label{3dcross}
         \vspace{15pt}
     \end{subfigure}
     \begin{subfigure}[t]{0.49\textwidth}
         \centering
         \includegraphics[scale=0.23]{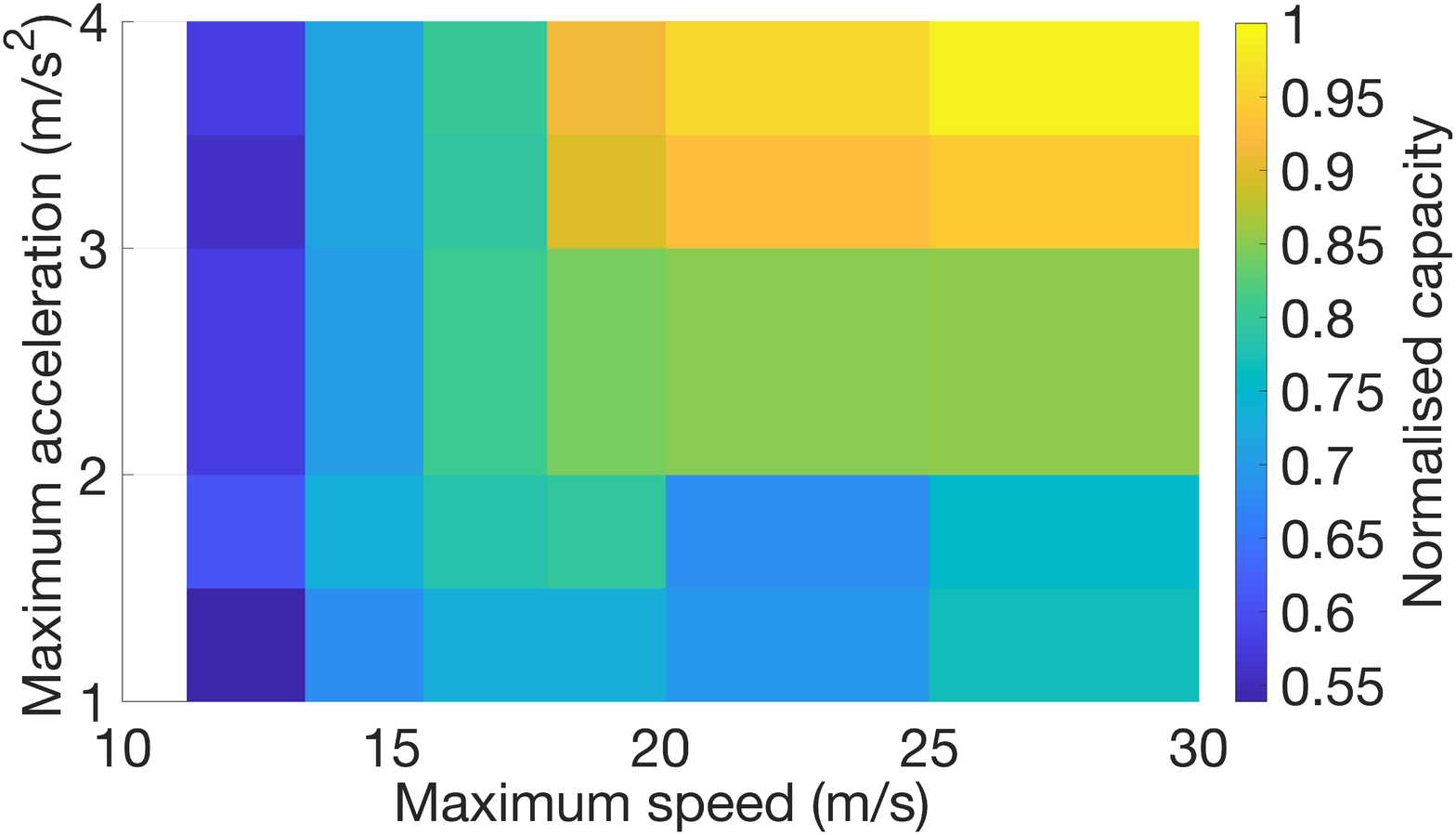}
         \caption{}
         \label{3dthrough}
    \end{subfigure}
    \vspace{-12pt}
    \caption{Sensitivity of the (a) crossing time and (b) capacity of the lane-free intersection in terms of the maximum speed and acceleration of CAVs. Initial speed of the vehicles is $10~(m/s$)}%
    \label{ModelTransformation}%
    \vspace{-10pt}
\end{figure*}

\begin{figure*}[t]
    \centering
    \begin{subfigure}[t]{0.49\textwidth}
         \centering
         \includegraphics[scale=0.22]{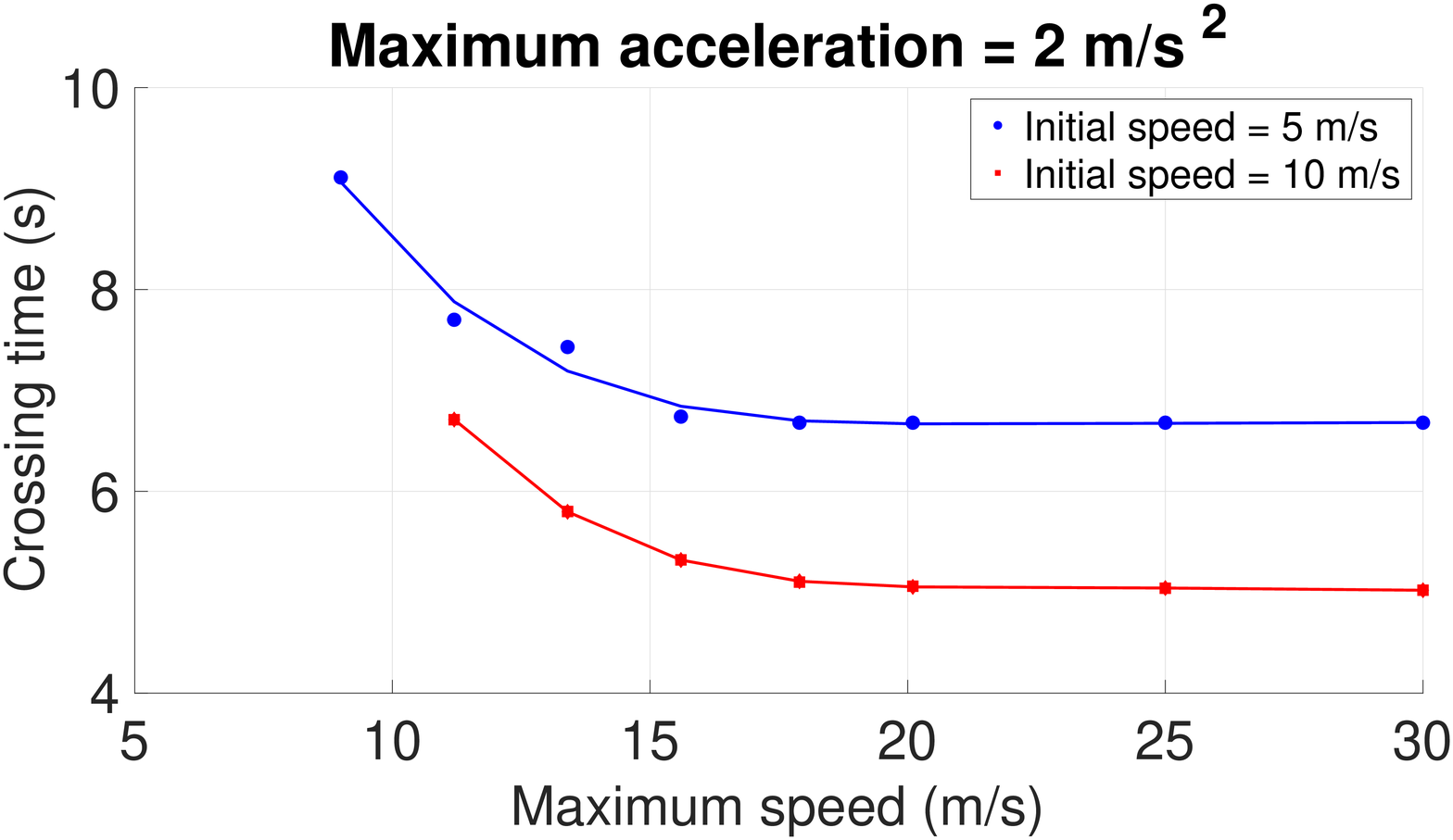}
         \caption{}
         \label{lanefree_sen_two_cross}
     \end{subfigure}
     \begin{subfigure}[t]{0.49\textwidth}
         \centering
         \includegraphics[scale=0.22]{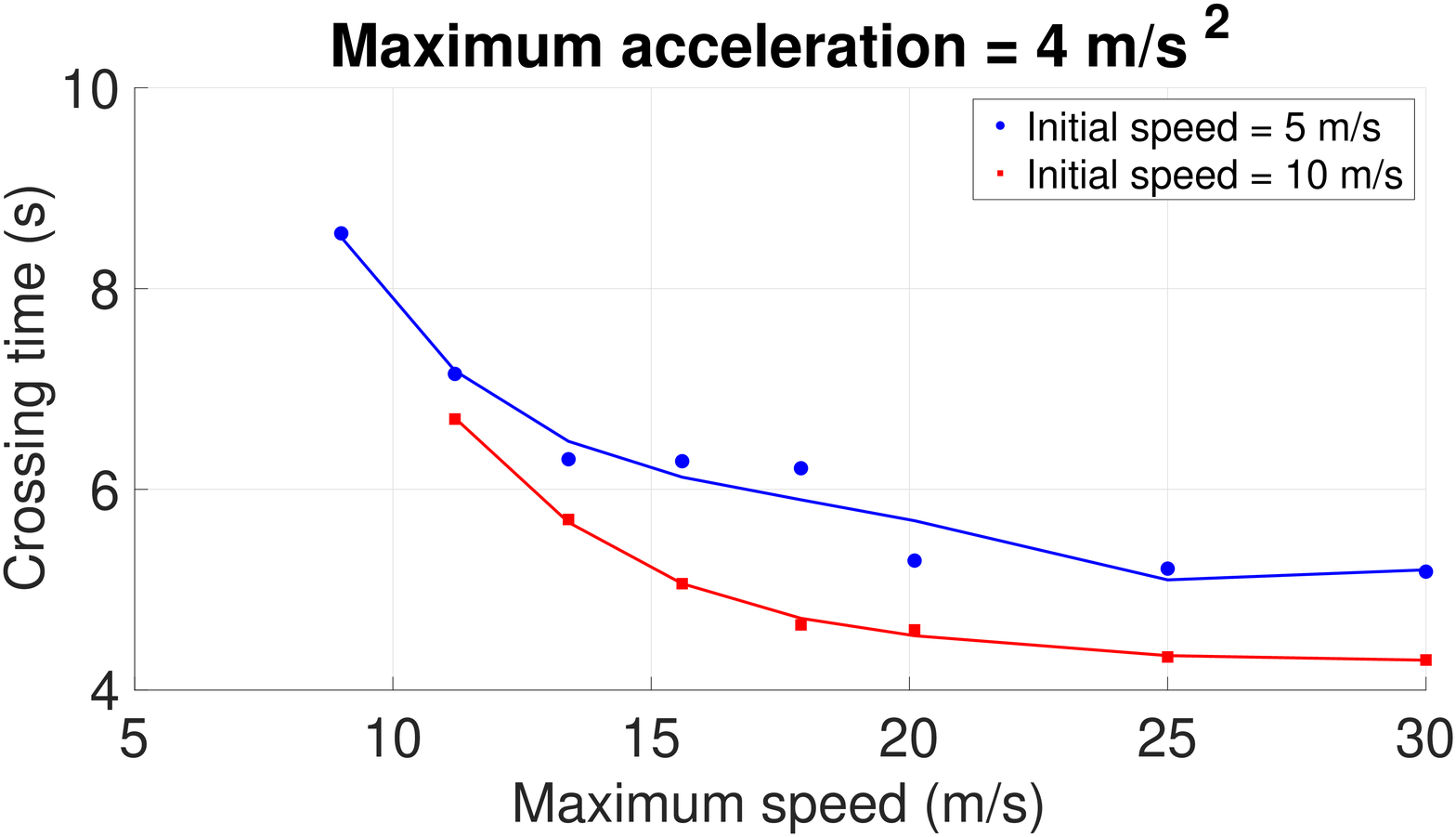}
         \caption{}
         \label{lanefree_sen_four_cross}
    \end{subfigure}
        \begin{subfigure}[t]{0.49\textwidth}
         \centering
         \includegraphics[scale=0.22]{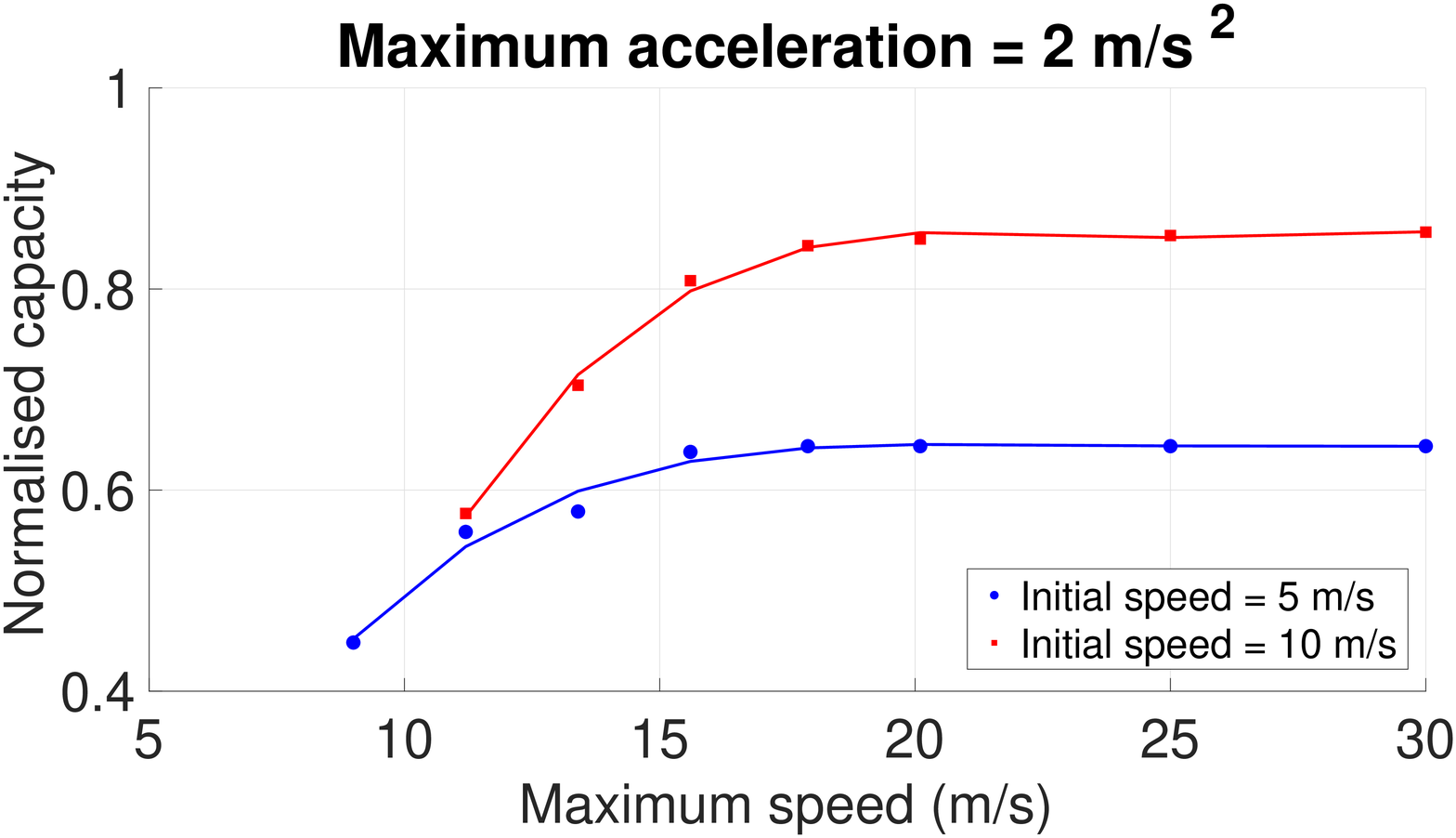}
         \caption{}
         \label{lanefree_sen_two_through}
         \vspace{15pt}
     \end{subfigure}
     \begin{subfigure}[t]{0.49\textwidth}
         \centering
         \includegraphics[scale=0.22]{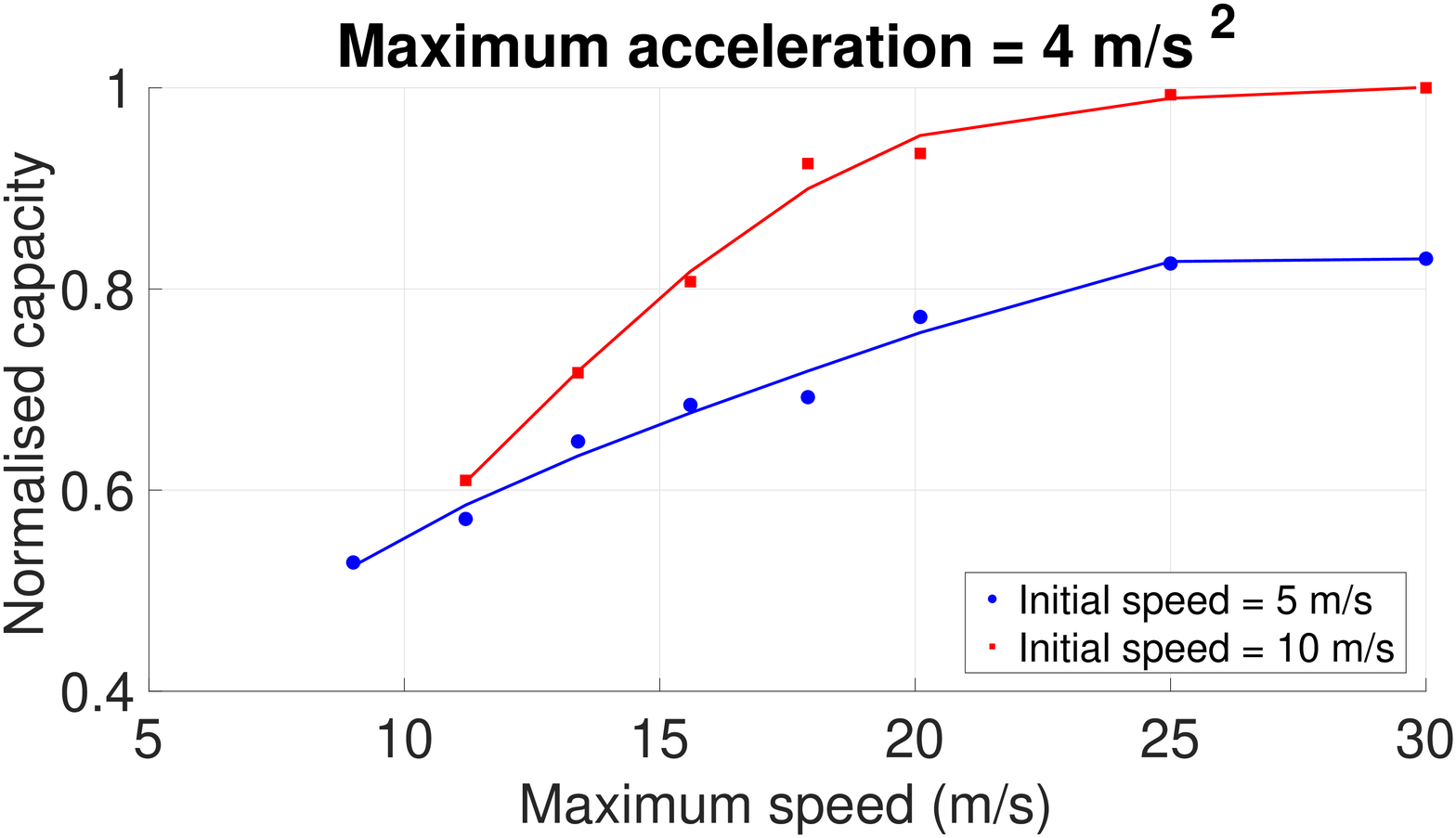}
         \caption{}
         \label{lanefree_sen_four_through}
    \end{subfigure}
    \vspace{-12pt}
    \caption{Crossing time and capacity of the lane-free intersection over different values of the maximum speed and for two values of the initial speed and maximum acceleration of CAVs. The solid lines are the corresponding fitted polynomials of order four which show the variation trends.}%
    \vspace{-10pt}\label{lane_sen}
\end{figure*}

Fig. \ref{num} shows the calculated throughput at different number of vehicles for the lane-free and signalised intersections. The throughput of both intersections increases until it reaches the capacity, as in Fig. \ref{num}, where the first collision happens in the lane-free case and throughput starts dropping in the signalised cases. It is worth noting that this different behaviour has no effect on the calculation of the maximum throughput. Fig. \ref{num} shows that the maximum number of crossing vehicles through the intersection is the same and equals 21 for both the lane-free and signalised with the max-pressure cases. However, it is slightly lower (i.e., 18 vehicles) for the signalised intersection with the Webster algorithm. 

Fig. \ref{num} also shows that capacity of the same intersection varies with different controlling algorithm. Particularly, the capacity of the intersection with a lane-free crossing of CAVs is $16,543~(veh/h)$ which shows an improvement of, respectively, 607\% and 743\% as compared to the signalised crossing with the max-pressure (with the capacity of $2,726\ (veh/h)$) and Webster (with the capacity of $2,227\ (veh/h)$) algorithms. This massive improvement is due to the facts that CAVs have shorter headway, do not stop by traffic lights and are able to use the most spatial-temporal area of the intersection. 

\section{Sensitivity Analysis of the Intersection Capacity}\label{sensitivity}
This section shows that capacity of the lane-free and signalised intersections varies by changes in the initial speed, maximum speed and maximum acceleration of vehicles. The crossing scenario is the same for both the intersections that makes the results comparable.

\subsection{Lane-free intersections}
Fig. \ref{3dcross} and Fig. \ref{3dthrough} show, respectively, variations of the crossing time and the normalised (with the maximum value) capacity of the lane-free intersection in Fig. \ref{Intersection_layout} due to changes in the maximum speed and acceleration of CAVs when the initial speed of the vehicles is $10~(m/s)$. The capacity is maximised and equivalently crossing time is minimised when both the maximum permissible speed of intersection and maximum acceleration are the largest possible that are, respectively, $30~(m/s)$ and $4~(m/s^2)$. 

\begin{figure*}[!t]
    \centering
    \begin{subfigure}[t]{0.49\textwidth}
         \centering
         \includegraphics[scale=0.22]{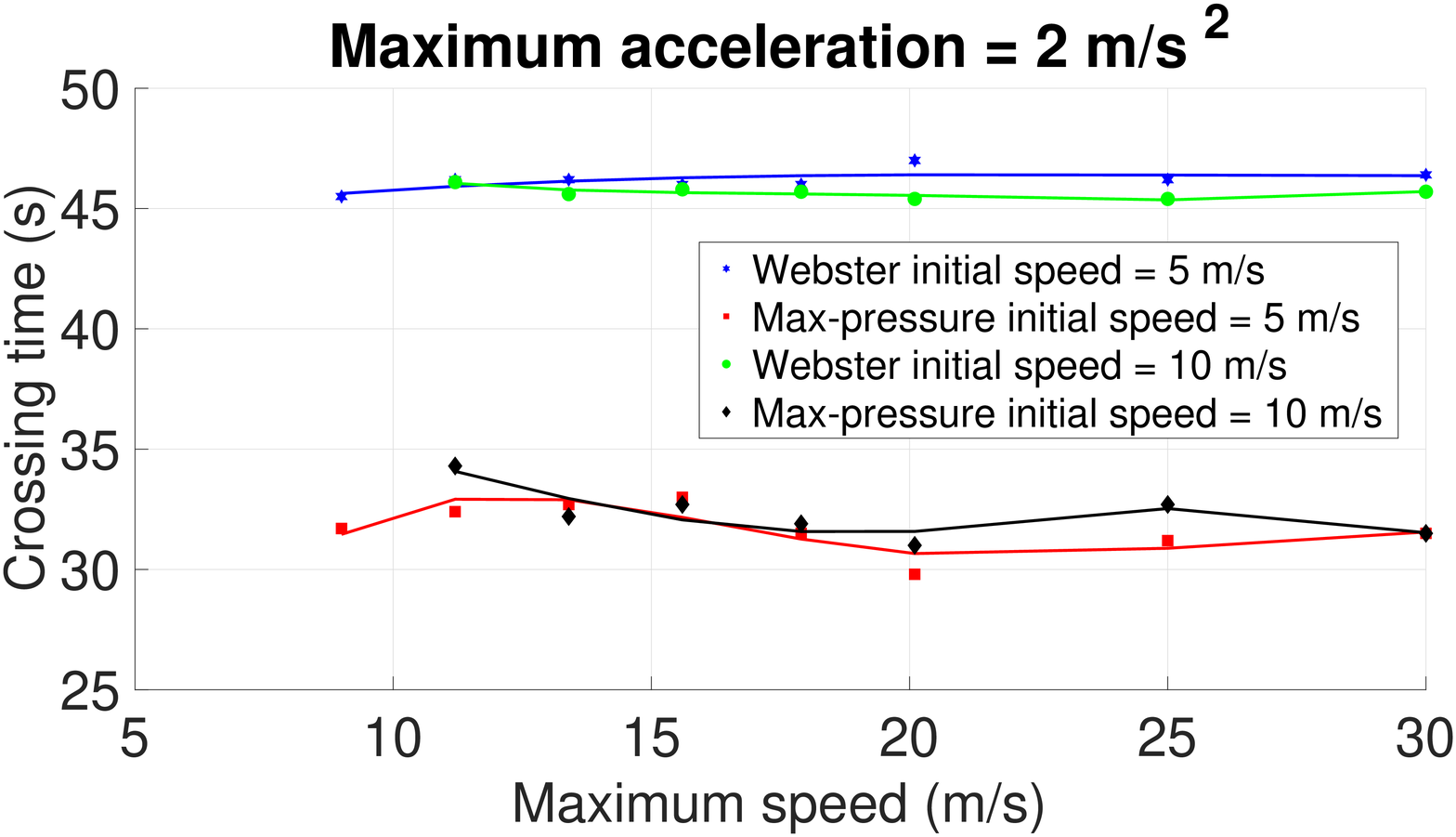}
         \caption{}
         \label{sig_sen_two_cross}
     \end{subfigure}
     \begin{subfigure}[t]{0.49\textwidth}
         \centering
         \includegraphics[scale=0.22]{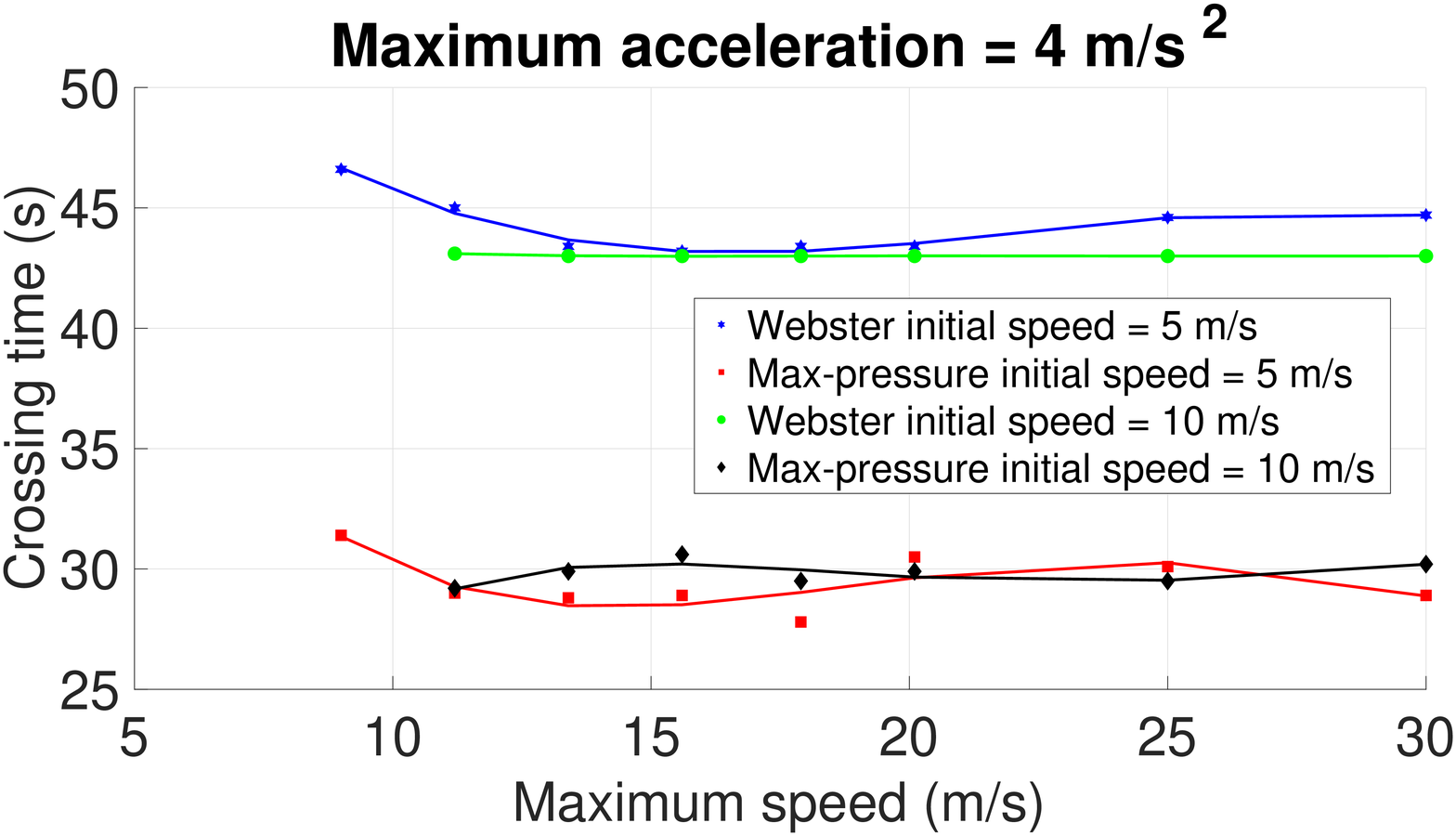}
         \caption{}
         \label{sig_sen_four_cross}
    \end{subfigure}
        \begin{subfigure}[t]{0.49\textwidth}
         \centering
         \includegraphics[scale=0.22]{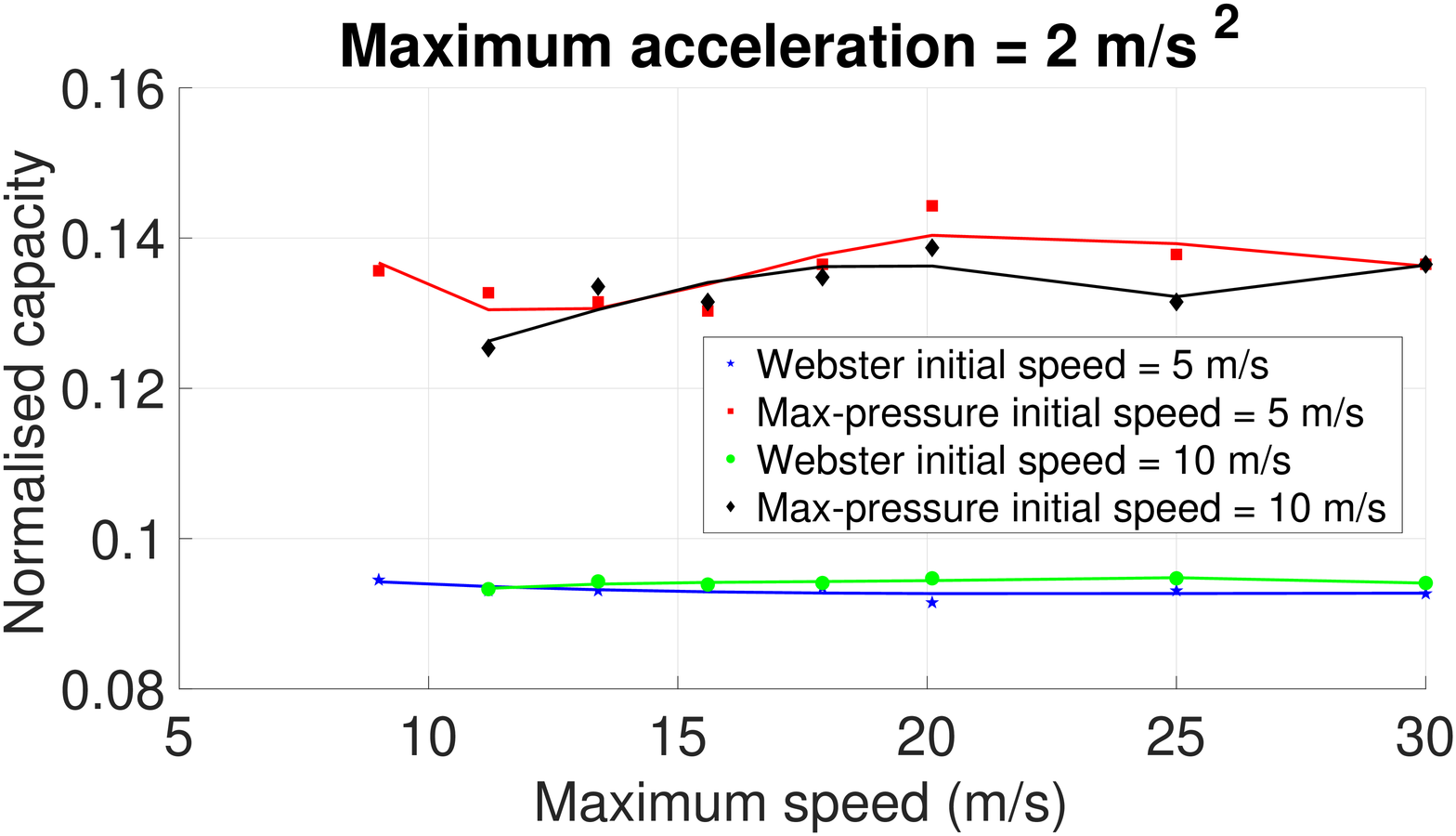}
         \caption{}
         \label{sig_sen_two_through}
     \end{subfigure}
     \begin{subfigure}[t]{0.49\textwidth}
         \centering
         \includegraphics[scale=0.22]{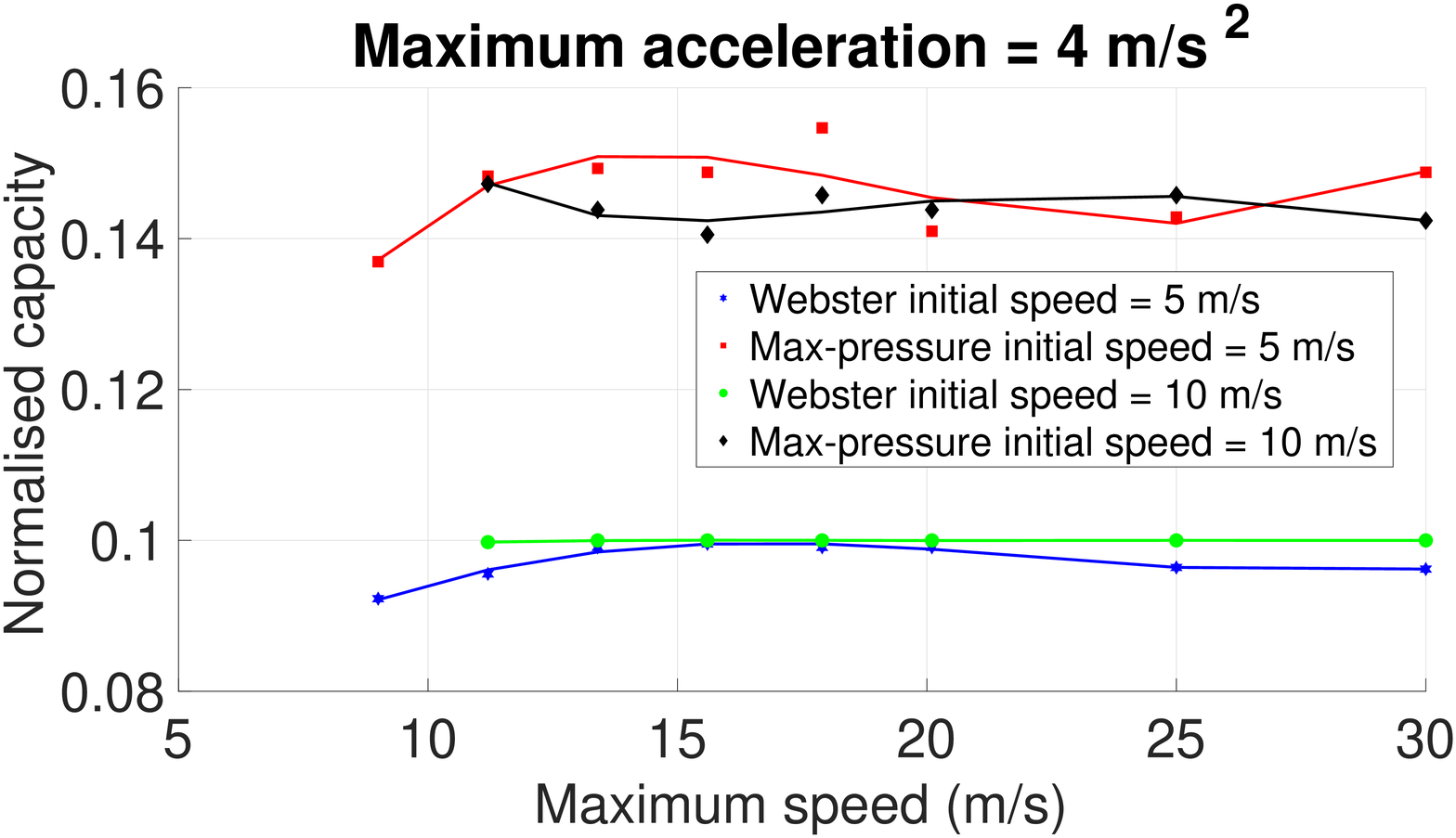}
         \caption{}
         \label{sig_sen_four_through}
    \end{subfigure}
    \caption{Capacity and crossing time of the signalised intersection for different values of the initial speed, maximum speed and maximum acceleration of the crossing HVs. The solid lines show trends of the variation as polynomials of order four.}%
    \label{sig_sen}
    \vspace{-10pt}
\end{figure*}

Fig. \ref{lane_sen} provides more detail of the results in Fig. \ref{ModelTransformation}. The results are illustrated by solid points which are fitted to a polynomial of order four with respect to the maximum speed. As seen, capacity of the lane-free intersection improves by, respectively, $28\%$ and $17\%$ due to an increase of the maximum acceleration from $2\ (m/s^2)$ to $4\ (m/s^2)$. Similarly, doubling the value of initial speed, the capacity increases by, respectively, $20\%$ and $15\%$ for the maximum accelerations of $2\ (m/s^2)$ and $4\ (m/s^2)$. 

Moreover, the crossing time and hence capacity of lane-free intersections are limited to a certain value which are specified by initial speed of the vehicles, as well as the geometry of the intersections. For example, Fig. \ref{lanefree_sen_two_cross} and Fig. \ref{lanefree_sen_four_cross} show that the crossing time is almost fixed to a constant value when the maximum speed of CAVs exceeds $18\ (m/s)$ and $25\ (m/s)$ for initial speed of, respectively, $5~(m/s)$ and $10~(m/s)$. In other words, even though CAVs are allowed to drive with a faster speed, yet they are not able to cross the intersection faster than a limit.

\subsection{Signalised intersections}
Fig. \ref{sig_sen} illustrates the variation of crossing time and equivalently capacity of the intersection in Fig. \ref{Intersection_layout} when there is a traffic light to control the flow of intersection with the max-pressure and Webster algorithms. Unlike capacity of the lane-free crossing of the intersection which, as shown in Fig. \ref{lane_sen}, strongly depends on initial speed, maximum speed and maximum acceleration of the crossing vehicles, capacity of the same intersection when is signalised with either of the algorithms is only slightly sensitive to the maximum permissible acceleration and does not vary by increasing the maximum or initial speeds of the crossing vehicles. This is shown in Fig. \ref{sig_sen_two_cross} and \ref{sig_sen_four_cross}. 

The major reason is that while CAVs cross the lane-free intersections continuously and with no interrupts, traffic lights oblige HVs to stop before the signalised intersections no matter what the vehicles' speed are. The crossing time $T$ of these stopped vehicles is dominated by the average human reaction time which is fixed to a constant value. Meanwhile, Fig. \ref{num} shows that the number of crossing vehicles $N$ at which throughput of the intersection reaches its capacity is almost fixed for a given intersection (e.g., that is 21 for the intersection in Fig. \ref{Intersection_layout}). Hence, referring to (\ref{capacity_proposed}), capacity of the signalised intersections is insensitive to the parameters.

\section{Conclusion}
\label{conclusion}
While it is known that the lane-free crossing of CAVs through intersections improves the capacity as compared to the signalised crossing of human drivers, to the best knowledge of the authors, there is no previous comprehensive analysis of the capacity of lane-free intersections. This is because the capacity measures which are used for the conventional intersections are not applicable to the lane-free crossing, and the crossing performance of CAVs depends on the collaborative behaviour of the vehicles and not the traffic light controller.

This work introduces a measure that consistently represents capacity of a given intersection for both the lane-free and signalised crossing, along with the algorithms to calculate the measure. The presented results show that the lane-free crossing of CAVs improves capacity of an intersection by 607\% and 743\% as compared to capacity of the signalised crossing of human drivers through the same intersection when traffic lights are controlled by, respectively, the max-pressure and Webster algorithms. The presented sensitivity analysis also shows that unlike the unresponsive capacity of the signalised crossing to the variation of initial speed, maximum crossing speed and maximum acceleration of vehicles, an increase with either of these parameters improves the performance of the lane-free crossing to a degree.

This work also provides a benchmark to evaluate the performance of the algorithms that will be developed to collaboratively cross CAVs through intersections. Future work will extend the provided analysis to the case with multiple intersections that consider more factors such as passenger comfort into the measurement of capacity.



\bibliographystyle{IEEEtran}

{\footnotesize
\bibliography{References}}

\vspace{-25pt}

\vfill

\end{document}